\documentclass[12pt]{article}

\usepackage{graphics,graphicx,cite,amssymb,epsfig,float,psfrag}
\usepackage[usenames]{color}
\usepackage{rotating}

\newcommand{\lsim}{\raisebox{-0.13cm}{~\shortstack{$<$ \\[-0.07cm] $\sim$}}~} 
\newcommand{\gsim}{\raisebox{-0.13cm}{~\shortstack{$>$ \\[-0.07cm] $\sim$}}~} 
 
\newcommand{\tb}{\tan\beta} 
\newcommand{\beq}{\begin{eqnarray}} 
\newcommand{\eeq}{\end{eqnarray}} 
\newcommand{\s}{\\ \vspace*{-4mm}}

\begin{document}
\thispagestyle{empty}

\vspace{1cm}

\hfill CERN--PH--TH/2011--310

\hfill LPT--ORSAY--11/122

\hfill LYCEN/2011--16

\vspace*{1.5cm}

\begin{center}

\mbox{\large\bf Implications of a 125 GeV Higgs for supersymmetric models}

\vspace*{.8cm}

A. Arbey$^{a,b,c}$, M. Battaglia$^{c,d,e}$, A.  Djouadi$^{c,f}$, F.
Mahmoudi$^{c,g}$ and J. Quevillon$^f$ 

\vspace*{8mm}

{\small 
    
$^a$ Universit\'e de Lyon, France; Universit\'e Lyon 1, CNRS/IN2P3, UMR5822 IPNL, 
F-69622~Villeurbanne Cedex, France.  

$^b$ Centre de Recherche Astrophysique de Lyon, Observatoire de Lyon, Saint-Genis Laval Cedex, F-69561, France; CNRS, UMR 5574; Ecole Normale Sup\'erieure de Lyon, Lyon, France,
France. 

$^c$ CERN, CH-1211 Geneva 23, Switzerland. 

$^d$ Santa Cruz Institute of Particle Physics, University of California, Santa Cruz,
CA 95064, USA. 

$^e$ Lawrence Berkeley National Laboratory, Berkeley, CA 94720, USA. 

$^f$ Laboratoire de Physique Th\'eorique, Universit\'e Paris XI and CNRS,
F--91405 Orsay, France.

$^g$ Clermont Universit\'e, Universit\'e Blaise Pascal, CNRS/IN2P3,\\LPC, BP 10448, 63000 Clermont-Ferrand, France. 

}
\end{center}

\vspace{.4cm}

\begin{abstract}  	

Preliminary results of the search for a Standard Model like Higgs boson at the LHC
with  5~fb$^{-1}$ data have just been presented  by the ATLAS and CMS collaborations
and an  excess of events at a mass of $\approx 125$ GeV has been reported.   If this
excess of events is confirmed by further searches with more data, it will have
extremely important consequences in the  context of  supersymmetric extensions of the
Standard Model and, in particular the minimal one, the MSSM. We show that for a
standard--like Higgs boson with a mass 123 $< M_h <$ 127 GeV, several unconstrained or
constrained  (i.e. with soft supersymmetry--breaking parameters unified at the high
scale) MSSM scenarios would  be excluded, while the parameters of some other scenarios 
would be severely restricted. Examples of constrained MSSM scenarios  which would be
disfavoured  as they predict a too light  Higgs particle are the minimal  anomaly and 
gauge mediated supersymmetry breaking models. The gravity mediated constrained MSSM
would still be viable,  provided the scalar top quarks are heavy and their trilinear
coupling large. Significant areas of the parameter space of models with heavy
supersymmetric  particles, such as split or high--scale supersymmetry, could also be
excluded as, in turn, they generally predict a too  heavy Higgs particle. 
\end{abstract} 

\newpage
\setcounter{page}{1}

\subsection*{1. Introduction} 

The ATLAS and CMS collaborations have released the preliminary results of their search
for the Standard Model (SM) Higgs boson at the LHC on almost  5~fb$^{-1}$ data per
experiment \cite{evidence}. While these results are not sufficient for the
two experiments to make any conclusive statement, the reported excess of events over
the SM background at a mass of $\sim$ 125~GeV offers a tantalising indication  that  the
first sign of the Higgs particle might be emerging. A Higgs particle with a mass of
$\approx 125$ GeV would be  a  triumph for the SM as the high--precision electroweak
data are hinting since many years to a light Higgs boson, $M_H \lsim 160$ GeV at the
95\% confidence level \cite{EW-data,PDG}. The ATLAS and CMS results, if confirmed,
would also have far reaching consequences for extensions of the SM and, in particular,
for supersymmetric theories (SUSY). The latter are widely considered to be the most
attractive extensions as they naturally protect the Higgs mass against large radiative
corrections and stabilise the hierarchy between the electroweak and Planck scales.
Furthermore, they allow for gauge coupling unification and the lightest SUSY particle
(LSP) is a good dark matter candidate; see Ref.~\cite{SUSY} for a review.\s

In the minimal SUSY extension, the Minimal Supersymmetric Standard Model (MSSM)
\cite{SUSY}, two Higgs doublet fields are required to break the electroweak symmetry,
leading to the existence of five Higgs particles: two CP--even $h$ and $H$, a CP--odd
$A$ and two charged $H^\pm$ particles \cite{Review}. Two parameters are needed to
describe the Higgs sector at the tree--level: one Higgs mass, which is generally taken
to be that of the pseudoscalar boson $M_A$, and the ratio of vacuum expectation values
of the two Higgs fields, $\tan\beta$, that is expected to lie in the range $1 \lsim \tb
\lsim 60$. At high $M_A$ values, $M_A \gg M_Z$, one is in the so--called decoupling
regime in which the neutral CP--even state $h$ is light and has almost exactly the
properties of the SM Higgs particle,  i.e. its couplings to fermions and gauge bosons
are the same,  while the  other CP--even state $H$ and the charged Higgs boson $H^\pm$
are heavy and degenerate in mass with the pseudoscalar Higgs particle, $M_H \approx
M_{H^\pm} \approx M_A$. In this regime, the Higgs sector of the MSSM thus looks almost
exactly as the one of  the SM with its unique Higgs particle. \s

There is, however,   one major difference between the two cases: while in the
SM  the Higgs mass is essentially a free parameter (and should simply be smaller
than about 1 TeV), the lightest CP--even Higgs particle in the MSSM is bounded
from above and, depending on the SUSY parameters that enter the radiative
corrections, it is restricted to values  \cite{Review,Sven}    
\beq  
M_h^{\rm max} \approx M_Z |\cos 2\beta|+{\rm radiative~corrections} \lsim
110\!-\! 135~{\rm GeV}   
\eeq 
Hence, the requirement that the $h$ boson mass coincides with the value of the 
Higgs particle ``observed" at the LHC, i.e. $M_h \approx 125$ GeV, would place
very strong constraints on the MSSM parameters through their contributions to
the radiative corrections to the Higgs sector.\s

In this paper, we discuss the consequences of such a value of $M_h$ for the MSSM. We 
first consider the unconstrained or the phenomenological MSSM  \cite{pMSSM} in which
the relevant soft SUSY--breaking parameters are allowed to vary freely (but with some
restrictions such as the absence of CP and flavour violation) and, then,  constrained
MSSM scenarios  (generically denoted by cMSSM here) such as the minimal supergravity
model (mSUGRA) \cite{mSUGRA}, gauge mediated (GMSB)  \cite{GMSB}  and  anomaly mediated
(AMSB) \cite{AMSB} supersymmetry breaking models. We also discuss the implications of
such an $M_h$ value for scenarios in which  the supersymmetric spectrum is extremely
heavy, the so--called split SUSY \cite{split} or high--scale SUSY \cite{heavy} models. 
\s

In the context of the phenomenological  MSSM, we  show that some scenarios which were
used as benchmarks for LEP2 and Tevatron Higgs analyses and are still used  at the LHC
~\cite{benchmarks} are excluded if $M_h \approx 125$ GeV, while some other  scenarios
are severely restricted. In particular, only when the SUSY--breaking scale is very
large and the mixing in the stop sector significant that one reaches this $M_h$ value. 
We also show that some constrained models, such as the minimal versions of GMSB and 
AMSB, do not allow for a sufficiently large mass of the lighter Higgs boson  and would
be disfavoured if the requirement $M_h \approx 125$ GeV is imposed. This requirement
sets  also strong constraints on the basic parameters of the mSUGRA scenario and only
small areas of the parameter space would be still allowed; this is particularly true 
in  mSUGRA versions in which one sets restrictions on the trilinear coupling. Finally,
in  the case of split or high--scale SUSY models, the resulting Higgs mass is in 
general much larger than $M_h \approx 125$ GeV and energy scales above  approximately 
$10^5$--$10^{8}$ GeV, depending on the value of $\tan\beta$, would also be disfavoured.

\subsection*{2. Implications in the phenomenological MSSM} 

The  value of the lightest CP--even Higgs boson mass $M_h^{\rm max}$ should in
principle depend on all the soft SUSY--breaking parameters which enter the
radiative corrections \cite{Sven}. In an unconstrained MSSM, there is a large
number of such parameters but analyses can be performed in the so--called 
``phenomenological MSSM" (pMSSM)~\cite{pMSSM}, in which CP conservation, flavour diagonal sfermion mass and
coupling matrices and universality of the first and second generations are imposed. The pMSSM involves 22 free parameters in addition to those of the
SM: besides $\tb$ and $M_A$, these are the higgsino mass parameter $\mu$,  the
three gaugino mass parameters $M_1, M_2$ and $M_3$, the diagonal left-- and
right--handed sfermion mass parameters $m_{ {\tilde f}_{L,R}}$ (5 for the third
generation sfermions and 5 others for the first/second generation sfermions)
and   the trilinear sfermion couplings $A_f$ (3 for the third generation  and 3
others for the first/second generation sfermions). Fortunately, most of these
parameters have only a marginal impact on the MSSM Higgs masses and, besides  $\tb$ and $M_A$, two of them play a major role: the SUSY breaking
scale that is  given in terms of the two top squark masses as $M_S = \sqrt{
m_{\tilde t_1} m_{\tilde t_2}}$ and  the mixing parameter in the stop sector,
$X_t = A_t -\mu \cot\beta$.\s

\noindent The maximal value of the $h$  mass, $M_h^{\rm max}$ is then obtained
for the following choice of parameters:

$i)$ a decoupling regime with a heavy pseudoscalar Higgs boson, 
$M_A \sim \mathcal{O}$(TeV);

$ii)$ large values of the parameter $\tb$, $\tb \gsim 10$;

$iii$ heavy stops, i.e. large $M_S$ and we choose $M_S=3$ TeV as a
maximal value\footnote{This value for $M_S$ would lead to an ``acceptable"
fine--tuning and would correspond to squark  masses of about 3 TeV, which is
close to the maximal value at which these particles can be detected at the 14
TeV LHC.}; 

$iv)$ a stop trilinear coupling $X_t=\sqrt{6}M_S$, the so--called maximal 
mixing scenario \cite{benchmarks}.\s

An estimate of the upper bound can be obtained by adopting the maximal mixing
scenario of  Ref.~\cite{benchmarks}, which is often used as a benchmark scenario
in Higgs analyses. We choose however  to be conservative, scaling the relevant
soft SUSY--breaking parameters by a factor of three compared to
Ref.~\cite{benchmarks} and using the upper limit $\tb \sim 60$:
\begin{eqnarray}
{M_h}^{\rm max}_{\rm bench} \ : \ \begin{array}{c} 
\tb=60 \, , \ M_S=M_A=3~{\rm TeV} \, , \ A_t=A_b =\sqrt{6}\,M_S\,, \\
M_2  \simeq 2 \,M_1 = |\mu|=\frac15 M_S \, , \ M_3=0.8 \,M_S\,.
\label{pbenchmark}
\end{array}
\end{eqnarray}

For the following values  of the top quark pole mass, the $\overline{\rm MS}$ 
bottom quark mass, the  electroweak gauge boson masses as well as the
electromagnetic and strong coupling constants defined at the scale $M_Z$, 
including  their $1\sigma$  allowed range \cite{PDG},
 \beq
m_t=172.9 \pm 1, \ \bar m_b (\bar m_b)=4.19 \pm 0.02, \
M_Z=91.19 \pm 0.002, \ M_W=80.42 \pm 0.003~{\rm [in~GeV]}  \nonumber \\ 
\alpha (M_Z^2) =1/127.916 \pm 0.015, \ \alpha_s(M_Z^2) =0.1184 \pm 0.0014
\hspace*{3cm}
\label{SM-inputs} 
\eeq
we use the programs {\tt Suspect} \cite{suspect}  and {\tt Softsusy} 
\cite{softsusy}  which calculate the Higgs and superparticle spectrum in the 
MSSM including the most  up--to--date information (in particular, they implement
in a similar way  the full one--loop and the dominant two--loop corrections
in  the Higgs sector; see Ref.~\cite{adkps}). One obtains the maximal 
value of the lighter Higgs boson, $M_h^{\rm max} \simeq 134$ GeV for maximal 
mixing. Hence, if one assumes that the particle observed at the LHC is the lightest MSSM
Higgs boson $h$, there is a significant portion of the pMSSM parameter space
which could  match the observed mass of $M_h \approx 125$ GeV in this 
scenario. However, in this case either $\tan\beta$ or the SUSY scale
$M_S$ should be  much lower  than in Eq.~(\ref{pbenchmark}).\s

In contrast, in the scenarios of no--mixing  $A_t \approx A_b \approx 0$ and 
typical mixing $A_t \approx A_b \approx M_S$ (with all other parameters  left as
in Eq.~(\ref{pbenchmark}) above) that are also used as benchmarks
\cite{benchmarks},  one obtains much  smaller $M_h^{\rm max}$ values than
compared to maximal mixing, $M_h^{\rm max} \simeq  121$ GeV and $M_h^{\rm max}
\simeq 125$ for, respectively, no--mixing and typical mixing.  Thus, if
$M_h\approx 125$ GeV, the  no--mixing scenario is entirely ruled out, while only
a small fraction of the  typical-mixing  scenario parameter space, with high
$\tan\beta$ and $M_S$  values, would survive. \s

The mass bounds above are  not yet fully optimised and  $M_h^{\rm max}$ values 
that are larger by  a few (1 or 2) GeV can be obtained by varying in a  reasonable range
the SUSY parameters entering the radiative corrections and  add an estimated
theoretical uncertainty\footnote{The theoretical  uncertainties in the
determination of $M_h$ should be small as the three--loop corrections to $M_h$
turn out  to be rather tiny, being less than 1 GeV \cite{3loop}.   Note that our
$M_h^{\rm max} $ values are slightly smaller  than the ones obtained in
Ref.~\cite{adkps} (despite of the higher $M_S$ used here) because of the
different top quark mass.} of about 1 GeV. To obtain a more precise
determination of $M_h^{\rm max}$ in the pMSSM,   we have again  used the 
programs {\tt Softsusy} and {\tt Suspect} to perform a flat scan of the pMSSM
parameter space by allowing its 22 input parameters to vary in an uncorrelated
way in the  following domains:
\begin{eqnarray}
1\leq \tb \leq 60 \, , \  50~{\rm GeV} \leq M_A \leq 3~{\rm TeV}\, , 
\ -9~{\rm TeV} \leq A_f \leq 9~{\rm TeV} \; ,   \nonumber \\
~~~50~{\rm GeV} \leq m_{\tilde f_L}, m_{\tilde f_R}, M_3 
\leq 3~{\rm TeV}\, , \ 
50~{\rm GeV} \leq M_1, M_2, |\mu|  
\leq 1.5~{\rm TeV}.
\label{scan-range}
\end{eqnarray}

\begin{figure}[t!]
\begin{center}
\mbox{
\includegraphics[width=0.5\textwidth]{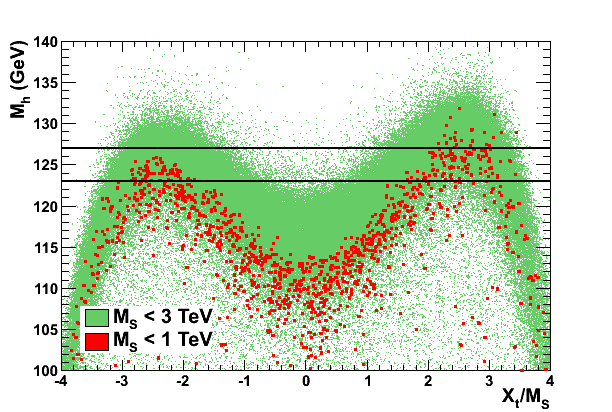}\hspace*{-5mm}
\includegraphics[width=0.5\textwidth]{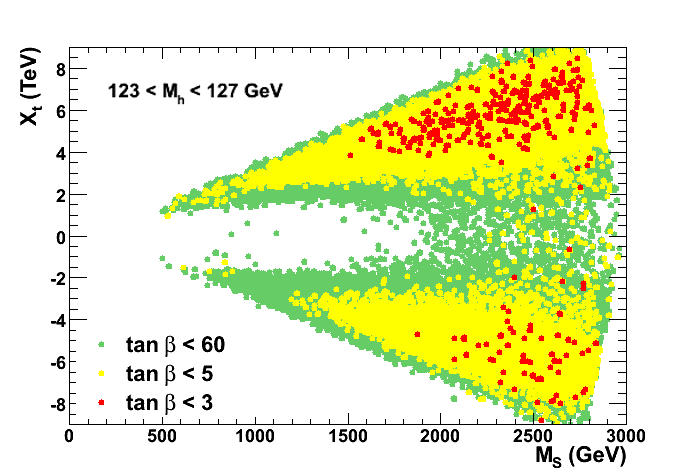}
}
\end{center}
\vspace*{-4mm}
\caption{\small The maximal value of the $h$ boson mass as a function 
of $X_t/M_S$ in the pMSSM when  all other soft SUSY--breaking parameters 
and $\tb$ are scanned in the range Eq.~(\ref{scan-range}) (left)
and the contours for 123$<M_h<$127 GeV in the $[M_S,X_t]$ plane for some
selected  range of $\tan\beta$ values (right).}
\label{Fig:pMSSM}
\vspace*{-3mm}
\end{figure}

We have discarded points in the parameter space that lead to a non--viable 
spectrum (such as charge and colour breaking minima which imposes the
constraint $A_t/M_s \lsim 3$) or to unrealistic Higgs masses (such as large
$\log (m_{\tilde g}/m_{\tilde t_{1,2}})$ terms that spoil the radiative 
corrections to $M_h$ \cite{adkps}). We select the Higgs
mass  for which 99\% of the scan points give a value  smaller than it.  The 
results are shown in Fig.~\ref{Fig:pMSSM} where, in the  left--hand side,  the
obtained maximal value of the  $h$ boson mass $M_h^{\rm max}$ is displayed as a
function of the ratio of  parameters $X_t/M_S$. The resulting values are
confronted to the mass range 
\beq  
123~{\rm GeV} \leq M_h \leq 127~{\rm GeV} \label{Mh-range}  
\eeq
where the upper limit corresponds to the 95\% confidence level bound reported by
the CMS collaboration \cite{evidence}, once the parametric uncertainties from the
SM inputs given in Eq.~(\ref{SM-inputs}) and the  theoretical uncertainties in
the determination of $M_h$ are included. Hence, only the scenarios with large
$X_t/M_S$ values and, in particular, those close to the maximal  mixing scenario
$A_t/M_S \approx \sqrt 6$ survive. The no--mixing scenario is ruled out for $M_S
\lsim 3$ TeV, while the typical mixing scenario  needs large $M_S$ and moderate
to large $\tan\beta$ values.  We obtain $M_h^{\rm max}$=136, 123 and 126 GeV in, the maximal, zero and typical mixing scenarios, respectively\footnote{We
have checked that  the program  {\tt FeynHiggs} \cite{feynhiggs} gives
comparable values for $M_h$  within $\approx 2$ GeV  which we  consider to be
our uncertainty as in Eq.~(\ref{Mh-range}).}.\s

The right--hand side of Fig.~\ref{Fig:pMSSM} shows the contours in the
$[M_S,X_t]$ plane where we obtain the mass range 123 GeV $< M_h <$ 127~GeV from
our pMSSM scan with $X_t/M_S \lsim 3$; the regions in which $\tan\beta \lsim 3,
5$ and 60 are highlighted. One sees again that a large part of the parameter
space is excluded  if the Higgs mass constraint is imposed\footnote{Note that
the $M_h^{\rm max}$ values given above are obtained with a heavy superparticle
spectrum, for which the constraints from flavour physics and sparticle searches
are evaded, and in the decoupling limit in which the $h$ production cross
sections and the decay branching ratios  are those of the SM Higgs boson.
However, we also searched for points in the parameter space in which the boson with mass $\simeq$ 125 GeV is the heavier CP--even $H^0$ boson which corresponds to values of $M_A$ of order 100 GeV. Among the $\approx 10^{6}$ valid MSSM points of the scan, only $\approx 1.5 \times 10^{-4}$ correspond to this
scenario. However, if we impose that the $H^0$ cross sections times branching
ratios are compatible with the SM values within a factor of 2 and include the
constraints from MSSM Higgs searches in the $\tau^+ \tau^-$ channel, only
$\approx 4 \times 10^{-5}$ of the points survive. These are all excluded once the $b\to
s\gamma$ and $ B_s \to \mu^+ \mu^-$ constraints are imposed. A detailed study of the
pMSSM Higgs sector including the dark matter and flavour constraints as well as
LHC Higgs and SUSY search limits is presented in Ref.~\cite{higgs-pmssm}.}. 

\subsection*{3. Implications for constrained MSSM scenarios} 

In constrained MSSM scenarios (cMSSM)\footnote{In this paper cMSSM denotes all
constrained MSSM scenarios, including GMSB and AMSB.},  the various soft SUSY--breaking
parameters obey a number of universal boundary conditions at a high energy scale such
as the GUT scale, thus reducing the number of basic input parameters to a handful. 
These inputs are evolved via the MSSM renormalisation group equations down to the low 
energy scale $M_S$ where the conditions of proper electroweak symmetry breaking (EWSB)
are imposed. The Higgs and superparticle spectrum is calculated, including the
important radiative corrections. Three classes of such models have been widely
discussed in the literature: 

-- The minimal supergravity (mSUGRA) model \cite{mSUGRA}, in which  SUSY--breaking is assumed to occur in a hidden sector
which communicates with the visible sector only via flavour-blind gravitational
interactions, leading to universal soft breaking terms. Besides the scale
$M_{\rm GUT}$ which is derived from the unification of the three gauge coupling
constants, mSUGRA has only four free parameters plus the sign of $\mu$:
$\tan\beta$ defined at the EWSB scale and $m_0, m_{1/2}, A_0$ which are
respectively, the common soft terms of all scalar masses, gaugino  masses and
trilinear scalar interactions, all defined at $M_{\rm GUT}$.\s

-- The gauge mediated SUSY--breaking (GMSB) model \cite{GMSB} in which
SUSY--breaking is communicated to the visible sector via gauge interactions. The
basic parameters of the minimal model are, besides $\tan\beta$ and sign$(\mu)$,
the messenger field  mass scale $M_{\rm mess}$, the  number of SU(5)
representations of the  messenger fields $N_{\rm mess}$ and the SUSY--breaking
scale in the visible sector $\Lambda$. To that, one adds the mass of the
LSP gravitino which does not play any role here. \s

-- The anomaly mediated SUSY--breaking (AMSB) model \cite{AMSB} in which 
SUSY--breaking is communicated to the visible sector via a  super-Weyl anomaly.
In the minimal AMSB version,  there are three basic parameters in addition  to
sign($\mu)$: $\tan\beta$, a universal parameter $m_0$ that  contributes to the
scalar masses at the GUT scale and the gravitino mass $m_{3/2}$.\s 

In the case of the mSUGRA scenario, we will in fact study four special 
cases:\s 

-- The no-scale scenario  with the requirement $m_0 \approx A_0 \approx 0$
\cite{no-scale}. This model leads  to  a viable spectrum compatible with all
present experimental constraints and with light staus for moderate $m_{1/2}$ 
and sufficiently high  $\tan\beta$ values; the mass of the gravitino (the
lightest SUSY particle) is a free parameter and can be adjusted  to provide the
right amount of dark matter.\s

-- A model with $m_0 \approx 0$ and $A_0 \approx -\frac14 m_{1/2}$ which,
approximately, corresponds  to the constrained next-to--MSSM (cNMSSM)
\cite{cNMSSM} in which a singlet Higgs superfield is added to the two doublet
superfields of the MSSM, whose components  however  mostly decouple from the
rest of the spectrum.  In this model, the requirement of a good singlino dark
matter candidate imposes $\tan\beta \gg 1$  and  the only relevant free
parameter is thus $m_{1/2}$ \cite{cNMSSM}.\s

-- A model with $A_0\! \approx\! -m_0$ which corresponds to a very constrained 
MSSM (VCMSSM)  similar to the one discussed in Ref.~\cite{new-bench} for input
values of the $B_0$ parameter close to zero.\s

-- The non--universal Higgs mass model (NUHM) in which the universal soft
SUSY--breaking scalar mass terms are  different for the sfermions and for the
two Higgs  doublet fields \cite{NUHM}. We will work in the general  case in
which, besides the four mSUGRA basic continuous inputs, there are two additional
parameters\footnote{ This scenario  corresponds to the  NUHM2 discussed
e.g. in Ref.~\cite{new-bench}; the model NUHM1 also discussed in
Refs.~\cite{NUHM,new-bench} and which has only one additional parameter is
simply a  special case of our NUHM scenario.} which can be taken to be $M_A$ and
$\mu$.  \s

In contrast to the pMSSM,  the various parameters which enter the radiative
corrections to the MSSM Higgs sector  are not all independent  in constrained 
scenarios as a consequence of the relations between SUSY breaking parameters
that are set at the high--energy scale and  the requirement that electroweak
symmetry breaking is triggered radiatively for each set of input parameters 
which leads to additional constraints. Hence, it is not possible to freely tune
the relevant weak--scale parameters to obtain the maximal value of $M_h$ given
previously. In order to obtain a reliable determination of 
$M_h^{\rm max}$  in a given constrained SUSY scenario, it is necessary to scan
through the allowed range of values for all relevant SUSY parameters.\s  

Following  the analysis performed in Ref.~\cite{adkps}, we  adopt  the  ranges
for the input parameters of the considered mSUGRA, GMSB and AMSB  scenarios: 

\begin{center} 
\begin{tabular}{rccc} 
mSUGRA: & 50 GeV $\leq m_0 \leq$ 3 TeV, & 50 GeV $\leq m_{1/2} \leq$ 3 TeV, & 
$|A_0|\leq 9 $ TeV; \\ 
GMSB: & 10 TeV $\leq \Lambda\leq$ 1000 TeV, & 1 $\leq M_{\rm mess} / \Lambda
\leq 10^{11}$, & $N_{\rm mess} =$ 1; \\ 
AMSB: & 1 TeV $ \leq m_{3/2}\leq$ 100 TeV, & 50 GeV $\leq m_0\leq$ 3 TeV. &  
\end{tabular}
\end{center}

Moreover, in the three cases we allow for both signs of $\mu$, require  $1\leq\tan
\beta\leq 60$ and, to avoid the need for excessive fine--tuning in the EWSB
conditions,  impose an additional bound on the weak--scale parameters, i.e.  $M_S =
M_{\rm EWSB} = \sqrt{m_{\tilde{t}_1}m_{\tilde{t}_2}} < 3~{\rm TeV}$.\s 

\begin{figure}[!h]
\begin{center}
\vspace*{-3mm}
\includegraphics[width=0.62\textwidth]{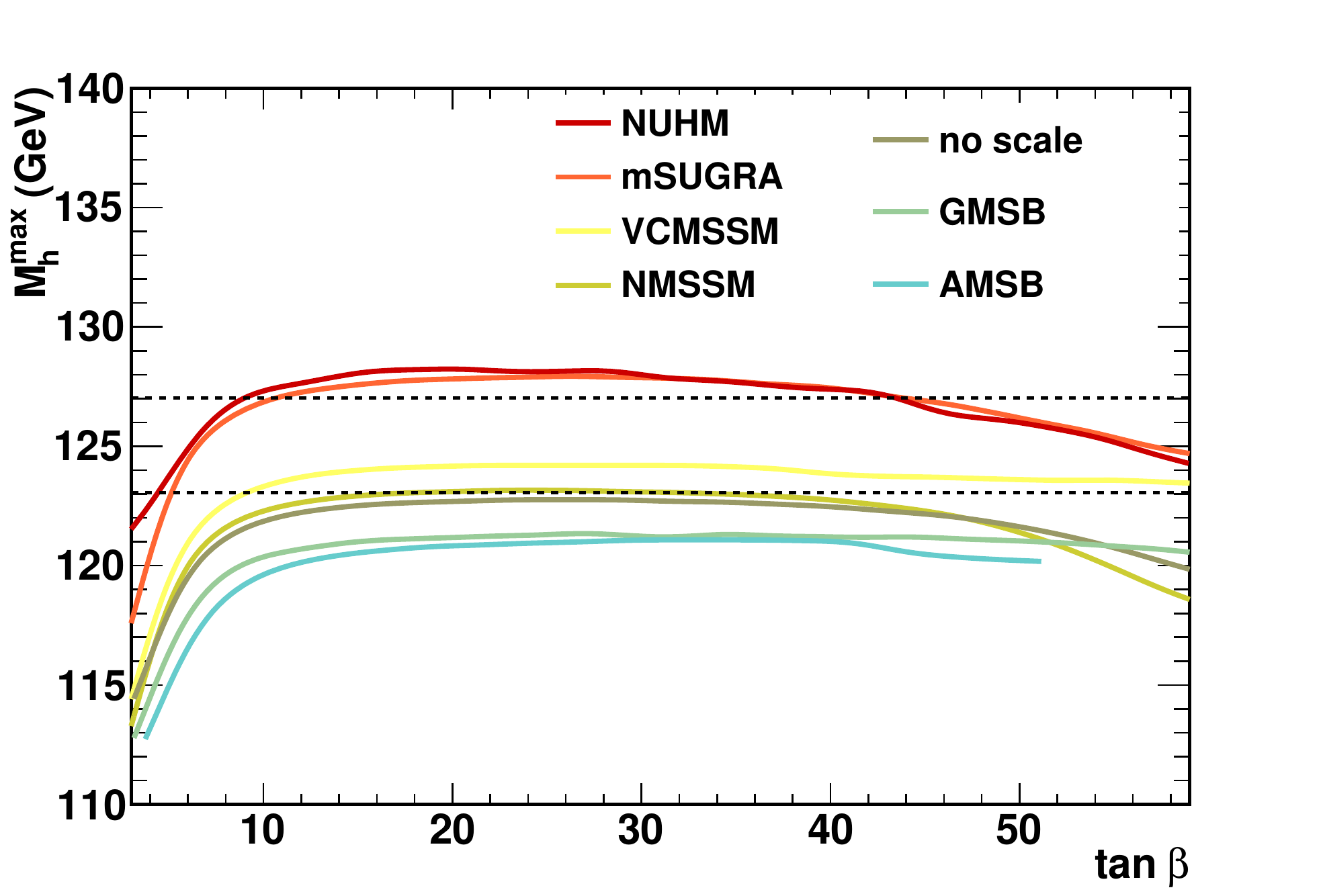}
\end{center}
\vspace*{-6mm}
\caption[]{\small The maximal value of the $h$ mass defined as the value for 
which 99\% of the scan points have a mass smaller than it, shown as a function 
of $\tan\beta$ for the various constrained MSSM models.}
\vspace*{-2mm}
\label{Fig:cMSSM}
\end{figure}

Using the programs {\tt Softsusy} and {\tt Suspect}, we have performed a  full
scan  of the GMSB, AMSB and mSUGRA scenarios, including the four options 
``no-scale'', ``cNMSSM'', ``VCMSSM'' and ``NUHM'' in  the later case. Using the
SM inputs of Eq.~(\ref{SM-inputs}) and varying the basic SUSY parameters of the
various models in the ranges described above, we have determined the maximal
$M_h$ value in each  scenario. The results for $M_h^{\rm max}$ are shown in
Fig.~\ref{Fig:cMSSM}  as a function of $\tan\beta$, the input parameter that is
common to all models. The highest $M_h$ values, defined as that   which have
99\% of the scan points below it, for any $\tan\beta$ value, are summarised in
Table~\ref{tab:mhmax}; one needs to add $\approx 1$ GeV to take into account the uncertainties in the SM inputs Eq.~(\ref{SM-inputs}). 

\begin{table}[!t]
\renewcommand{\arraystretch}{1.4}
\begin{center}
\begin{tabular}{|l|c|c|c||c|c|c|c|}\hline
model & AMSB & GMSB & mSUGRA & no-scale & cNMSSM & VCMSSM& NUHM \\ \hline
$M_h^{\rm max}$ & 121.0 & 121.5 & 128.0 & 123.0 & 123.5 & 124.5 & 128.5 \\ \hline
\end{tabular}
\end{center}
\vspace*{-0.5cm}
\caption{Maximal $h^0$ boson mass (in GeV) in the various constrained 
MSSM scenarios 
 when scanning over all the input parameters in the ranges described
in the text.}
\label{tab:mhmax}
\end{table}

In all cases, the maximal $M_h$ value is obtained for $\tan \beta$ around 20. 
We observe that in the adopted parameter space of the models and with the
central values of the SM inputs, the upper $h$ mass value (rounded to the upper
half GeV) is $M_h^{\rm max} = $ 121 GeV in AMSB, i.e. much less that 125
GeV, while in the  GMSB scenario one has $M_h^{\rm max} = $ 121.5 GeV. Thus,
clearly, the two scenarios are disfavoured if the lightest CP--even Higgs
particle has indeed a mass in the range 123~$< M_h <$~127~GeV. In the case of
mSUGRA, we obtain a maximal value  $M_h^{\rm max} = 128$ GeV and, thus, some
parameter space of the model would still survive the $M_h$ constraint.

\begin{figure}[!t]
\begin{center}
\mbox{
\includegraphics[width=0.5\textwidth]{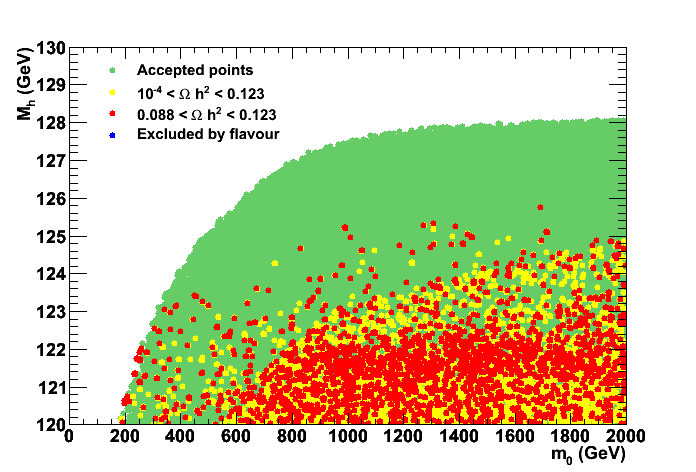}\hspace*{-5mm}
\includegraphics[width=0.5\textwidth]{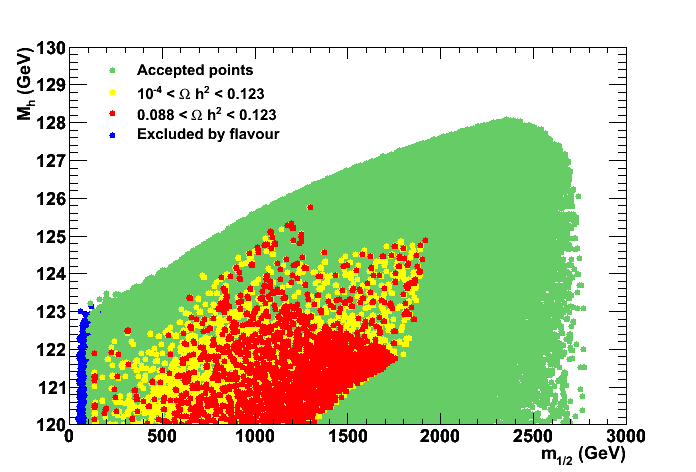}
}\\
\mbox{
\includegraphics[width=0.5\textwidth]{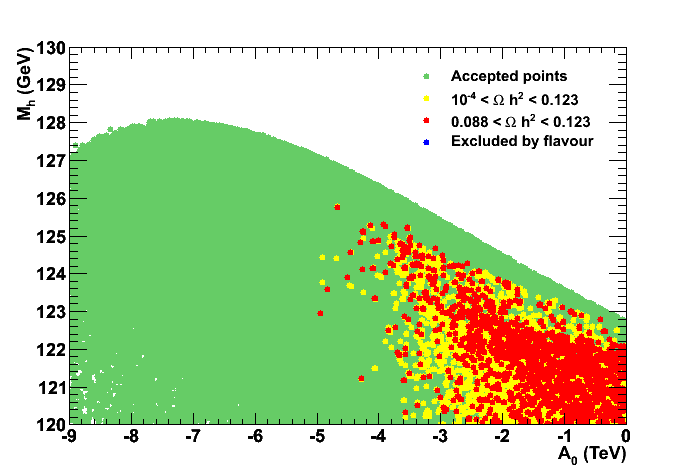}\hspace*{-5mm}
\includegraphics[width=0.5\textwidth]{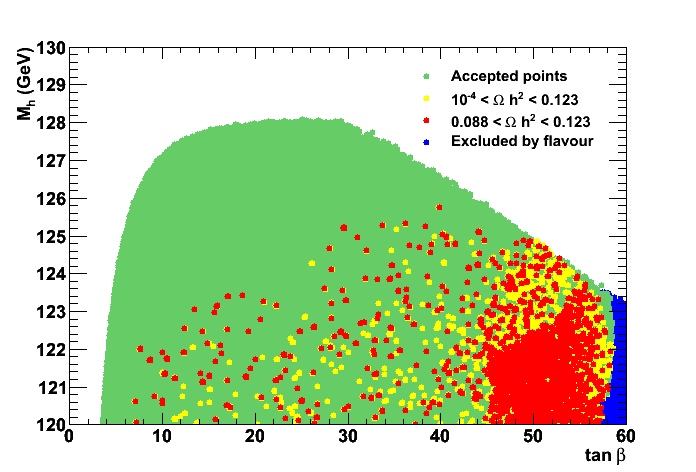}
}
\end{center}
\vspace*{-5mm}
\caption[]{\small The value of $M_h$ as a function of  one mSUGRA continuous 
parameter when a scan is  performed on the other parameters. The 
constraints from Higgs and SUSY searches  at the LHC are included and the 
impact of flavour ($b\to s\gamma, B_s \to \mu^+
\mu^-$, $B\to \tau\nu$) and DM constraints are shown.} 
\vspace*{-2mm}
\label{Fig:mSUGRA}
\end{figure}

\begin{figure}[!t]
\begin{center}
\mbox{
\includegraphics[width=0.52\textwidth]{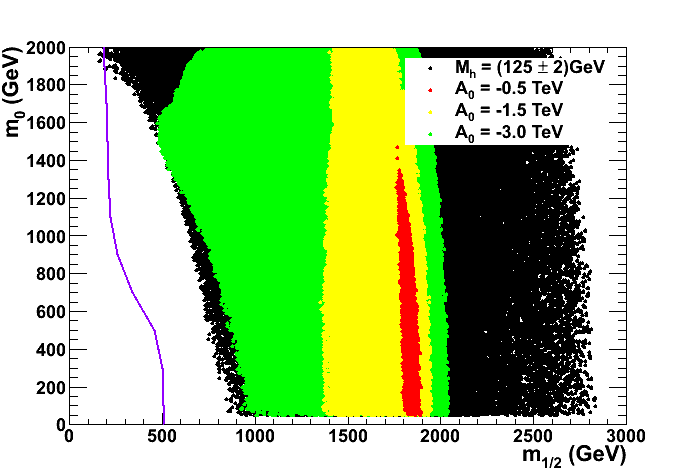}\hspace*{-5mm}
\includegraphics[width=0.52\textwidth]{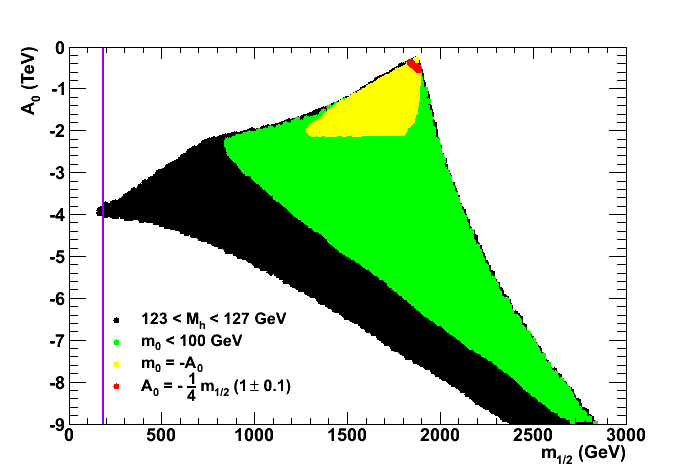}
}
\end{center}
\vspace*{-5mm}
\caption[]{\small Contours in which 123 $< M_h <$ 127 GeV, resulting of a full scan 
of the mSUGRA parameter but for particular choices of the inputs $A_0$ (left) and 
$m_0$ (right). The lower bound from LHC searches of SUSY strongly interacting particles 
in the fully hadronic channel with 1 fb$^{-1}$ data \cite{CMS-SUSY} is shown by a continuous line.} 
\vspace*{-2mm}
\label{Fig:mSUGRA2}
\end{figure}

The upper bound on $M_h$ in these scenarios  can be qualitatively understood by
considering in each model the allowed values of the trilinear coupling $A_t$,
which essentially determines the stop mixing parameter $X_t$ and thus the value
of $M_h$ for a given scale $M_S$. In GMSB, one has $A_t\approx 0$ at relatively
low scales and its magnitude does not significantly increase in the evolution
down to the scale  $M_S$; this implies that we are almost in the no--mixing
scenario which  gives a low value of $M_h$ as can be seen from
Fig.~\ref{Fig:pMSSM}. In  AMSB, one has a non-zero $A_t$ that is fully predicted
at any renormalisation scale in terms of the Yukawa and gauge couplings;
however, the ratio $A_t/M_S$ with $M_S$ determined  from the overall SUSY
breaking scale $m_{3/2}$ turns out to be rather small, implying again that we
are close to the no--mixing scenario.  Finally, in the mSUGRA model, since we
have allowed $A_t$ to vary in a wide range as $|A_0|\leq  9$ TeV,  one can get a
large $A_t/M_S$ ratio which leads to a heavier Higgs particle. However, one
cannot easily reach $A_t$ values such that $X_t/M_S \approx \sqrt 6$ so that we
are not in the maximal--mixing scenario  and  the higher upper bound on $M_h$ in
the pMSSM is not reached.\s

In turn, in two particular cases of mSUGRA that we have discussed in addition,
the ``no--scale" and the  ``approximate cNMSSM" scenarios, the upper bound on
$M_h$ is much lower than in the more general mSUGRA case and, in fact, barely
reaches the value $M_h \approx 123$ GeV. The main reason is that these
scenarios  involve small values of $A_0$ at the GUT scale, $A_0 \approx 0$ for
no--scale and $A_0 \approx -\frac14 m_{1/2}$ for the cNMSSM. One then obtains
$A_t$ values at the weak scale that are too low to generate a significant stop
mixing and, hence, one is again close to the no--mixing scenario. Thus, only a
very small fraction of the parameter space of these two sub--classes of the
mSUGRA model survive (in fact, those leading to the  $M_h^{\rm max}$ value)
if we impose 123 $< M_h <$ 127 GeV. These models hence should have a very
heavy spectrum as a value $M_S \gsim 3$ TeV is required to increase $M_h^{\rm
max}$. In the VCMSSM, $M_h \simeq 124.5$ GeV can be reached as $|A_0|$
can be large for large $m_0$, $A_0 \approx -m_0$, allowing at least for typical
mixing.  \s

Finally, since the NUHM  is more general than  mSUGRA as we have two more free
parameters, the $[\tan\beta, M_h]$ area shown in  Fig.~\ref{Fig:cMSSM} is larger
than in the mSUGRA case. However, since we are in the decoupling regime and the
value of $M_A$ does not matter much (as long as it a larger than a few hundred
GeV) and  the key weak--scale parameters entering the determination of $M_h$,
i.e. $\tan \beta, M_S$ and $A_t$ are approximately the same in both models, one
obtains a bound $M_h^{\rm max}$ that is only slightly higher in NUHM compared to
mSUGRA. Thus, the same discussion above on the mSUGRA scenario, holds also true
in the NUHM case. \s

In the case of the ``general" mSUGRA model, we show in Figs.~\ref{Fig:mSUGRA}
and \ref{Fig:mSUGRA2} some contours in the parameter space which highlight some
of the points discussed above. Following Ref.~\cite{scans} where the relevant
details can be found, constraints\footnote{All the points in Fig.~4 correspond
to the decoupling regime of the MSSM Higgs sector and, hence, to an $h$ boson
with SM cross sections and branching ratios. Furthermore,  as the resulting SUSY
spectrum for $M_h\!=\!125\!\pm \!2$ GeV is rather heavy in  constrained
scenarios, one obtains very small contributions to  $(g-2)_\mu$.} from the LHC
in Higgs  \cite{higgs-pmssm} and superparticle searches  \cite{CMS-SUSY}  and
the  measurement of $B_s \to \mu^+\mu^-$ as well as the requirement of a correct
cosmological density as required by WMAP have been implemented. We use the
program {\tt SuperIso Relic} \cite{superiso} for the calculation of dark matter relic density
and flavour constraints.

\subsection*{4. Split and high--scale SUSY models} 

In the preceding discussion, we have always assumed that the SUSY--breaking
scale is  relatively low, $M_S \lsim 3$ TeV, which implies that some of the 
supersymmetric and heavier Higgs particles could be observed at the LHC or at
some other TeV collider. However, as already mentioned,  this choice is mainly
dictated by fine--tuning considerations which are a rather subjective matter as
there is no compelling criterion to quantify the acceptable amount of tuning.
One could well have a very large value of $M_S$ which implies  that, except for
the lightest $h$ boson, no other scalar particle is accessible at the LHC or at
any foreseen collider.\s

This  argument has been advocated to construct the so--called split SUSY
scenario \cite{split} in which the soft SUSY--breaking mass terms for all the
scalars of the theory, except for one Higgs doublet, are extremely large, i.e. 
their common value $M_S$ is such that $M_S \gg 1$ TeV (such a situation occurs
e.g. in some string motivated models, see Ref.~\cite{heavy-string}). Instead,
the mass parameters for the spin--$\frac12$ particles,  the gauginos and the
higgsinos, are left in the vicinity of the EWSB scale,  allowing for a solution
to the dark matter  problem and a successful gauge coupling unification, the two
other SUSY virtues. The split SUSY models are much more predictive than the usual
pMSSM as only a handful parameters are needed to describe the low energy theory.
Besides the common value $M_S$ of the soft SUSY-breaking sfermion and one Higgs
mass parameters, the basic inputs are essentially the three gaugino masses
$M_1,M_2,M_3$ (which can be unified to a common value at $M_{\rm GUT}$ as in
mSUGRA), the higgsino parameter $\mu$ and $\tan\beta$. The  trilinear couplings
$A_f$, which are expected to have values close to the EWSB scale, and thus  much
smaller than $M_S$, will in general play a  negligible role. \s

Concerning the Higgs sector, the main feature of split SUSY is that at the 
high scale $M_S$,  the boundary condition on the quartic Higgs coupling  of 
the theory is determined by SUSY:
\beq
\label{boundlam}
\lambda(M_S) = \frac{1}{4}\left[ g^2(M_S)+g^{\prime 2}(M_S)
\right] \,\cos^22\beta~. 
\eeq
where $g$ and $g'$ are the SU(2) and U(1) gauge couplings. Here, $\tan\beta$ is
not a parameter of the low-energy effective theory: it enters only the boundary
condition above and cannot be interpreted as the ratio of two Higgs vacuum
expectation values. In this case, it should not be assumed to be larger than 
unity as usual and will indeed adopt the choice  $1/60 \leq \tan\beta \leq 60$. 

If the scalars are very heavy, they will lead to radiative corrections in the
Higgs sector that are significantly  enhanced by large logarithms, $\log( 
M_{\rm EWSB}/M_S)$, where $M_{\rm EWSB}$ is the scale set by the gaugino and
higgsino masses. In order to have reliable predictions, one has to properly
decouple the heavy states from the low-energy theory and resum the large
logarithmic corrections; in addition, the radiative corrections due to the
gauginos and the higgsinos  have to be implemented.  Following the early work of
Ref.~\cite{split}, a comprehensive study of the split SUSY spectrum has  been
performed in Ref.~\cite{bds}; see also Ref.~\cite{gian} that appeared recently.
All the features of the model have been implemented in the Fortran code {\tt
SuSpect} upon which the numerical analysis presented here is based.\s 

One can adopt an even more radical attitude than in the split SUSY case and
assume that the gauginos and higgsinos are also very heavy, with a mass close to
the scale $M_S$; this is the case in the so--called high--scale SUSY model
\cite{heavy}.  Here, one abandons  not only the SUSY solution to the fine-tuning
problem but also the solution to the dark matter problem by means of  the  LSP
and the successful unification of the gauge coupling constants.  However, there
will still be a trace of SUSY at low energy: the matching of the SUSY and the
low--energy theories is indeed encoded in the Higgs quartic coupling  $\lambda$
given by Eq.~(\ref{boundlam}). Hence, even if broken at very high scales, SUSY
would still lead to a ``light" Higgs boson whose mass will contain information
on $M_S$ and $\tan\beta$. \s

\begin{figure}[!t]
\vspace*{-1mm}
\begin{center}
\hspace*{-2mm}
\mbox{
\includegraphics[width=0.51\textwidth]{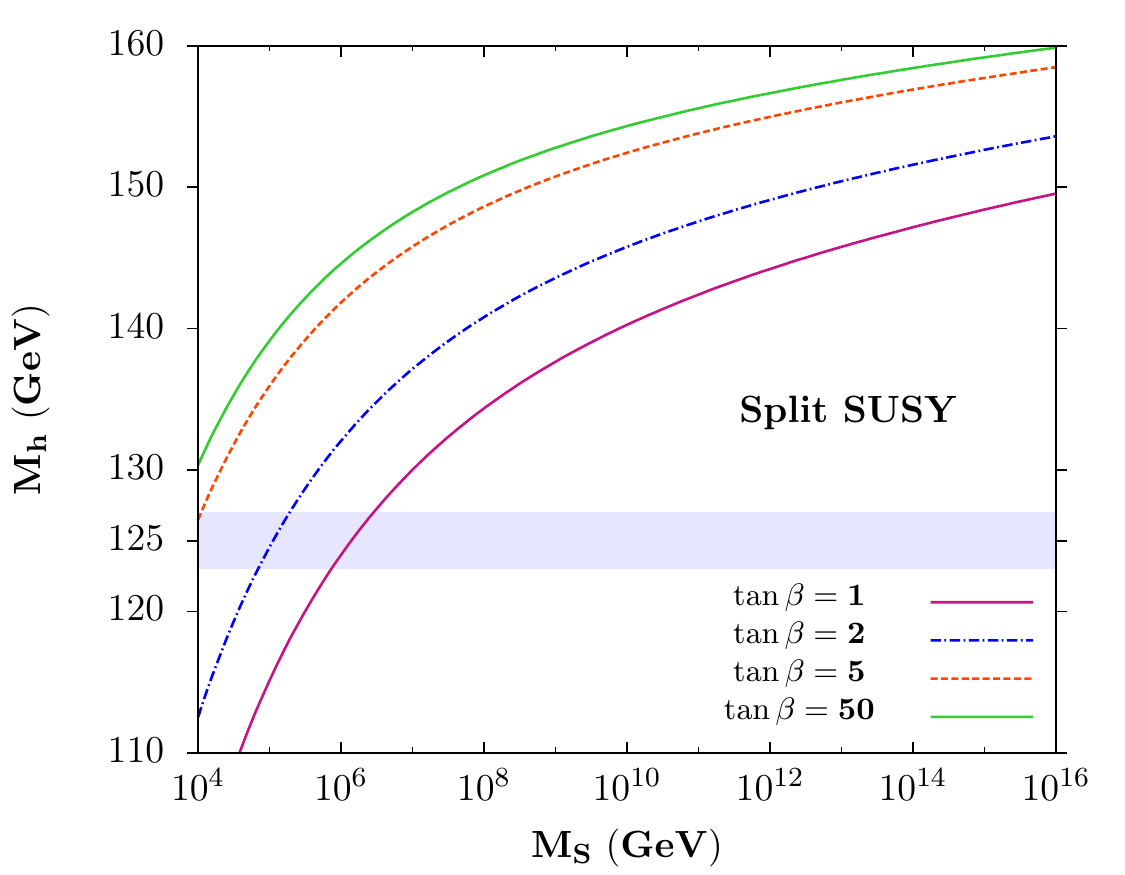}\hspace{-3mm} 
\includegraphics[width=0.51\textwidth]{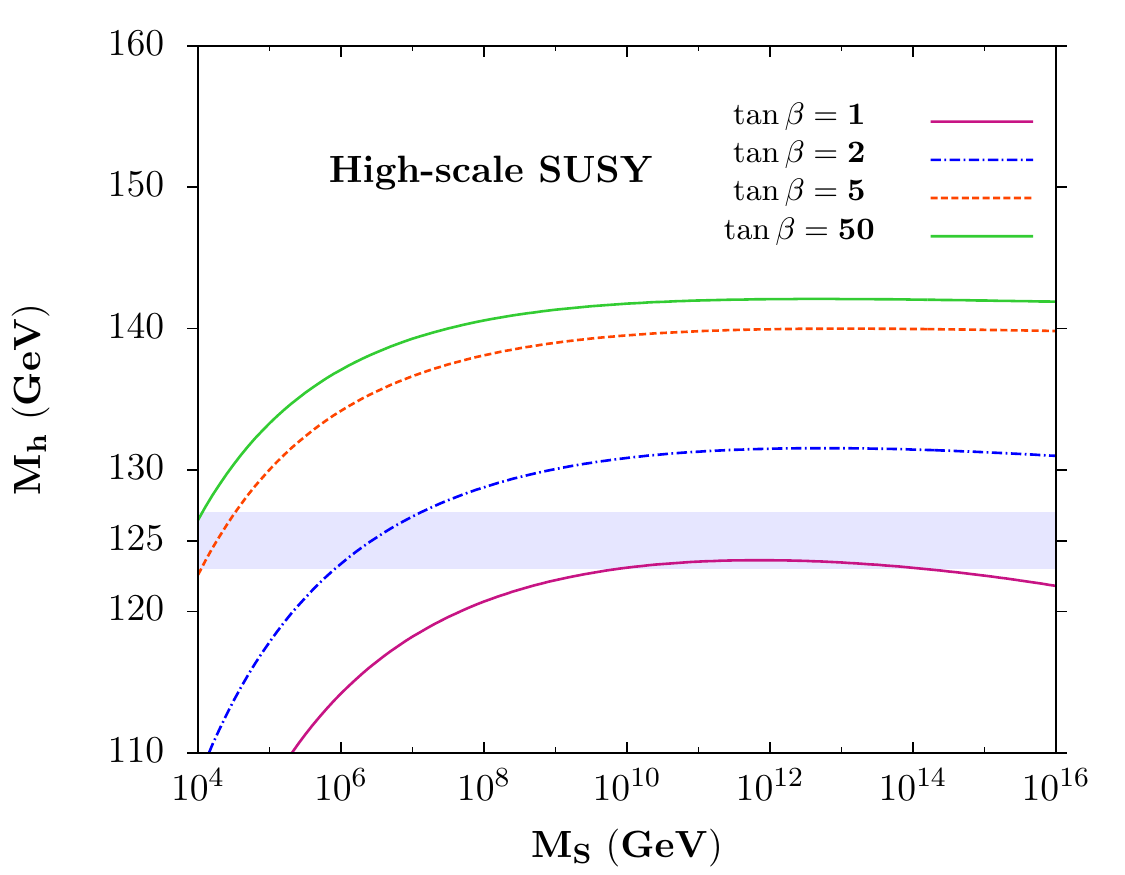}
}
\end{center}
\vspace*{-7mm}
\caption[]{\small The value of $M_h$ as a function of $M_S$ for several values 
of $\tan\beta=1,2,5,50$ in the split SUSY (left) and high--scale SUSY (right) 
scenarios.}
\vspace*{-2mm}
\label{Fig:hMSSM}
\end{figure}

The treatment of the Higgs sector of the high--scale SUSY scenario is similar to
that of split SUSY: one simply needs to decouple the gauginos and higgsinos
from the low energy spectrum (in particular remove their contributions to the
renormalisation group evolution of the gauge and Yukawa couplings and to the
radiative corrections to the $h$ boson mass) and set their masses to $M_S$. We
have adapted the version of the program {\tt Suspect} which handles the 
split SUSY case to also cover the case where $M_1\approx  M_2 \approx M_3
\approx |\mu| \approx M_S$. Using this program, we have  performed a
scan in the $[\tan\beta, M_S]$ plane  to determine the value of $M_h$ in the
split SUSY and high--scale SUSY scenarios. The values given in
Eq.~(\ref{SM-inputs}) for the SM input parameters  have been adopted and, in the
case of split SUSY, we have chosen $M_{\rm EWSB} \approx \sqrt{|M_2\mu| }
\approx 246$ GeV for the low scale.  The results are shown in
Fig.~\ref{Fig:hMSSM}. In this figure $M_h$ is displayed as a function of
$M_S$ for selected values of $\tan\beta$ in split and heavy--scale SUSY.\s

As expected, the maximal $M_h$ values are obtained at high $\tan\beta$ and $M_S$
values and, at the scale $M_S \approx 10^{16}$ GeV at which the couplings $g$
and $g'$ approximately unify in the split SUSY scenario, one obtains $M_h
\approx 160$ GeV for the higher $\tan\beta=50$ value\footnote{ Our result is
different by a few GeV from that given in Ref.~\cite{gian} as  the
gaugino/higgsino  two loop RGEs were used in that reference while we include
only the one--loop RGEs,  and different choices for scales have been adopted.
This points to  sizable theoretical uncertainties that we are presently
analysing.}. We do not include the error bands in the  SM inputs which would
lead to an uncertainty of about 2 GeV on $M_h$, mainly due to the 1 GeV
uncertainty on the top quark mass. In addition, we have assumed the zero--mixing
scenario as the parameter $A_t$ is expected to be much smaller than $M_S$; this
approximation might not be valid for $M_S$ values below 10 TeV and a maximal
mixing $A_t/M_S = \sqrt 6$ would increase the Higgs mass value by up to 10 GeV
at $M_S ={\cal O} (1~{\rm TeV})$ as was discussed earlier for the pMSSM. In the
high--scale SUSY scenario, we obtain a value  $M_h\approx 142$ GeV  (with
again an uncertainty of approximately 2 GeV from the top mass) for high
$\tan\beta$ values and at the unification scale $M_S \approx 10^{14}$ GeV as in 
Ref.~\cite{gian,heavy}. Much smaller $M_h$ values, in the 120 GeV range, can be
obtained for lower scales and $\tan\beta$. \s

Hence, the requirement that the Higgs boson mass is in the range 123 $< M_h <$ 127 GeV
imposes strong constraints on the parameters of these two models. For this
Higgs mass range, very large scales are needed for $\tan\beta\approx 1$ in the 
split (high--scale) SUSY scenario, while scales not too far from $M_S\!
\approx \! 10^{4}~{\rm GeV}$ are required at high $\tan\beta$. Thus,  even
in these extreme scenarios,  SUSY should manifest itself at scales much below
$M_{\rm GUT}$ if  $M_h\approx 125$ GeV. 

\subsection*{5. Conclusions}

We have discussed the impact of a Standard Model--like Higgs boson with a mass
$M_h \approx 125$ GeV on supersymmetric theories in the context of both
unconstrained and constrained MSSM scenarios. We have shown that in the
phenomenological MSSM, strong restrictions can be set on the mixing in the top
sector and, for instance,  the no--mixing scenario is excluded unless the
supersymmetry breaking scale is extremely large, $M_S \gg 1$ TeV, while the
maximal mixing scenario is disfavoured for large $M_S$ and $\tb$ values.\s

In constrained MSSM scenarios, the impact is even stronger. Several scenarios,
such as minimal AMSB and GMSB are  disfavoured as they lead to a too light $h$
particle. In the mSUGRA case, including the possibility that the Higgs mass
parameters  are non--universal,  the allowed part of the parameter space should
have large stop masses and $A_0$ values. In more constrained versions of this
model such as the ``no--scale" and approximate ``cNMSSM" scenarios, only a very 
small portion of the parameter space is allowed by the Higgs mass bound.\s

Finally, significant areas of the parameter space of models with large $M_S$
values leading to very heavy supersymmetric  particles, such as split SUSY or
high--scale SUSY, can also be excluded as, in turn, they tend to predict a too 
heavy Higgs particle with $M_h \gsim 125$ GeV.\bigskip

\noindent {\bf Acknowledgements:} We thank Pietro Slavich for discussions.

\end{document}